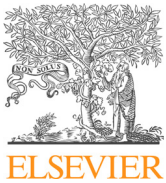
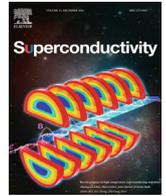

Review Article

# Recent progress in high-temperature superconducting undulators

Zhuangwei Chen [a,b,c], Marco Calvi [d], John Durrell [e], Cristian Boffo [f], Dabin Wei [a,b,c], Kai Zhang [c,*], Zhentang Zhao [g,*]

[a] *Shanghai Institute of Applied Physics, Chinese Academy of Sciences, Shanghai 201800, China*
[b] *University of Chinese Academy of Sciences, Beijing 100049, China*
[c] *Zhangjiang Laboratory, Shanghai 201210, China*
[d] *Center for Photon Science, Paul Scherrer Institute, 5232 Villigen PSI, Switzerland*
[e] *Bulk Superconductivity Group, Department of Engineering, University of Cambridge, Cambridge CB21PZ, United Kingdom*
[f] *Fermi National Accelerator Laboratory, P.O. Box 500, Batavia, IL 60510, USA*
[g] *Shanghai Advanced Research Institute, Chinese Academy of Sciences, Shanghai 201210, China*

ABSTRACT

Considerable effort has been devoted to the development of superconducting undulators (SCUs) intended for particle accelerator-based light sources, including synchrotrons and free electron laser (FEL) facilities. Recently, a high-temperature superconducting (HTS) undulator prototype, consisting of staggered-array Re-Ba-Cu–O bulks, achieved an on-axis sinusoidal magnetic field profile with a peak amplitude $B_0$ of 2.1 T and a period length of 10 mm, resulting in a deflection parameter $K = 1.96$. Such a short period HTS undulator not only enables the generation of higher-energy photons, but also supports the construction of economically feasible and compact FELs with shorter linear accelerators (LINACs). This article provides a comprehensive review of recent advances in the staggered-array bulk HTS undulator as well as other types of HTS undulators. Furthermore, it offers insights into the development of engineering HTS undulator prototypes designed for deployment in synchrotron and free electron laser (FEL) facilities. We conclude by discussing opportunities for and the challenges facing the use of HTS undulators in practical applications.

## Contents

# 1. Introduction

Over recent decades, there has been a rising demand in synchrotrons and free electron lasers (FELs) for insertion devices with enhanced performance. One approach to meeting this demand involves the realisation of undulator technologies capable of generating high on-axis magnetic fields over short periods. Significant evolution in undulator designs has occurred, encompassing permanent magnet undulators (PMUs) [1], in-vacuum undulators (IVUs) [2], cryogenic permanent magnet undulators (CPMUs) [3], superconducting undulators (SCUs) [4,5], and most recently, high-temperature superconducting undulators (HTSUs) [6,7]. Undulators with shorter periods promise substantial improvements in photon energy output and could facilitate the development of next-generation compact FELs that require shorter linear accelerators (LINACs). A notable example is the International Linear Collider (ILC) project, which necessitates the deployment of 11.5 mm-period superconducting helical undulators. These undulators are crucial for generating circularly polarized $\gamma$-ray sources aimed at producing polarized positrons [8–10].

Superconducting undulators offer significant advantages over PMUs because they achieve higher on-axis magnetic field ($B_0$) generation for a given period ($\lambda_u$) [5], thereby enhancing photon flux in the high-energy region of the x-ray spectrum. The Karlsruhe Institute of Technology (KIT) and Argonne National Laboratory (ANL) have actively developed and implemented SCU technologies using Nb-Ti/Cu superconducting coils within their respective storage rings for beamline applications [11–16]. Currently, SCU installations at these facilities utilize $\lambda_u$ exceeding 14 mm to achieve satisfactory deflection parameters, $K = 0.934 B_0[T]\lambda_u[cm]$. This is primarily due to the requirement of accommodating electron bunches within a large-bore vacuum pipe while they circulate in a closed orbit (e.g., 7 mm for KIT synchrotron and 7.2 mm for APS). In recent years, the Shanghai HIgh repetitioN rate XFEL and Extreme light facility (SHINE) as well as the European XFEL (EuXFEL) in collaboration with Bilfinger Noel have been developing SCUs with reduced beam gap and increased undulator field $B_0$ [17–20]. These developments aim to support the construction of the 3rd hard x-ray beamline at SHINE and to serve as SCU afterburners at EuXFEL's SASE2 hard x-ray beamline [21,22].

Superconducting undulators utilizing $Nb_3Sn$/Cu wires are expected to achieve elevated undulator fields due to their superior critical current density ($J_c$) compared to Nb-Ti/Cu wires. Argonne National Laboratory (ANL) and Lawrence Berkeley National Laboratory (LBNL) have developed expertise in essential technologies necessary for the construction of $Nb_3Sn$ undulators. These include the wind-and-react technique and stress management strategies aimed at minimizing $J_c$ degradation in heat-treated $Nb_3Sn$ coils [23–27]. In 2023, ANL in collaboration with LBNL and Fermi National Accelerator Laboratory (FNAL) achieved a significant milestone by fabricating a 1.1 m-long $Nb_3Sn$ planar undulator, achieving an undulator field of $B_0 = 1.17$ T with a period of 18 mm and a magnetic gap of 9.5 mm. Subsequently, this device was successfully integrated into the storage ring of Advanced Photon Source (APS) at ANL and delivered effective operational performance for users before the shutdown of APS for upgrade [28].

Since 2018, Paul Scherrer Institute (PSI) has been dedicated to advancing high-temperature superconducting (HTS) undulators with period lengths as short as 10 mm [7,29–37]. Their short planar prototype, utilizing staggered-array Re-Ba-Cu–O (ReBCO) bulks, achieved a record undulator field of $B_{y0} = 2.1$ T at a period of $\lambda_u = 10$ mm, corresponding to a $K$ value of 1.96. Additionally, a short helical prototype employing double staggered-array ReBCO bulks achieved an on-axis field of $B_{x0} = B_{y0} = 2.6$ T at a period of $\lambda_u = 16$ mm [38]. These achievements represent significant strides in undulator technology. Both the Swiss Light Source 2.0 (SLS2.0) and the Shanghai Soft X-ray Free-Electron Laser (SXFEL) have outlined plans to develop practical prototypes of bulk HTS undulators for integration into the storage ring or downstream the LINACs [21,39,40].

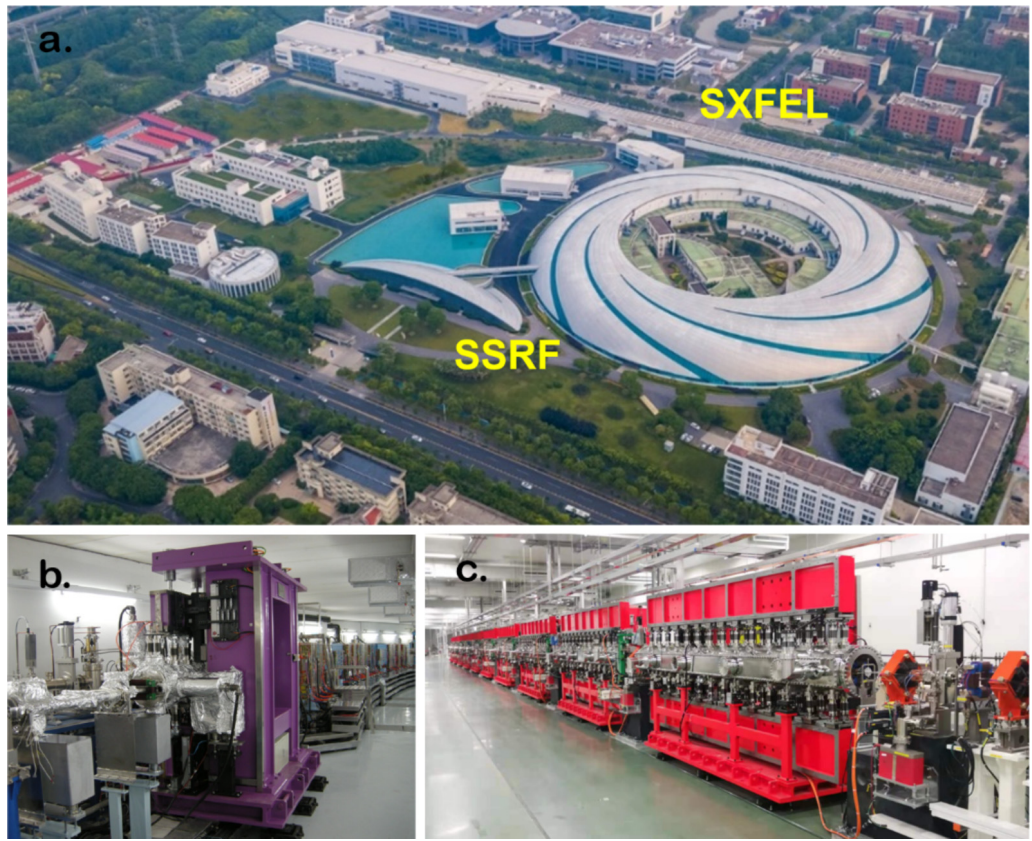

**Fig. 1.** (a) Aerial photograph of SSRF and SXFEL constructed in Zhangjiang Science City, Shanghai. (b) Undulator unit in the storage ring of SSRF. (c) Undulator system downstream the LINACs of SXFEL.

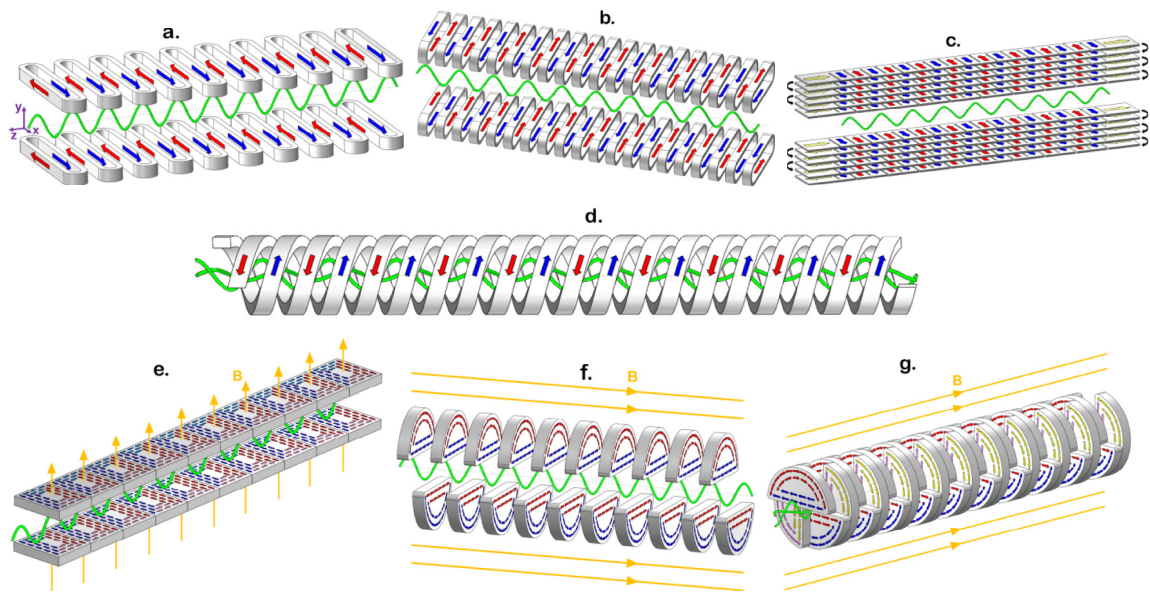

**Fig. 2.** HTS planar undulator comprising of two sets of **(a)** horizontal racetrack coils, **(b)** vertical racetrack coils, and **(c)** laser-structured, laminated ReBCO tapes. **(d)** HTS helical undulator comprising of one pair of helically wound superconducting coils. **(e)** HTS planar undulator comprising of two sets of horizontally arranged, magnetized ReBCO bulks. **(f)** HTS planar undulator comprising of staggered-array ReBCO bulks. **(g)** HTS helical undulator comprising of double staggered-array ReBCO bulks. Each type of HTS undulator model is created with ten periods. The green lines refer to the on-axis sinusoidal magnetic field $B_y(z)$ or $B_x(z)$. (For interpretation of the references to colour in this figure legend, the reader is referred to the web version of this article.)

Fig. 1(a) presents an aerial perspective of the Shanghai Synchrotron Radiation Facility (SSRF) and Shanghai Soft X-ray Free-Electron Laser (SXFEL) located within Zhangjiang Science City, Shanghai [41,42]. Undulators are positioned either within straight sections of the storage ring or downstream of the LINACs to produce highly collimated photon beams with a high degree of transverse and temporal coherence, as depicted in Fig. 1(b-c). Simultaneously, construction progresses on SHINE, a superconducting high-repetition-rate hard X-ray Free-Electron Laser (FEL) facility spanning approximately 3.1 km in length and capable of accelerating electron bunches up to 8 GeV [43,44].

This paper first presents an overview of the working principles of and recent advancements in various configurations of HTS undulators utilizing ReBCO bulk or coated conductors. It subsequently details the developmental progress of engineering prototypes of HTS undulators designated for the Swiss Light Source upgrade (SLS2.0) and the SXFEL facility [39,42]. Finally, the paper discusses opportunities and challenges of HTS undulators for practical applications.

## 2. Working principles

There are two primary configurations of HTS undulators: planar and helical. In planar undulators, high-energy electron bunches oscillate within either the *x-z* plane under the influence of an alternating magnetic field $B_y(z)$ as shown in Fig. 2(a), or within the *y-z* plane under the influence of an alternating magnetic field $B_x(z)$, oriented along the beam axis. This process results in the generation of linearly polarized photons either horizontally or vertically. In contrast, helical undulators use alternating magnetic fields in the *x-z* and *y-z* planes, with a 90-degree phase shift between them, causing electron bunches to follow a spiral trajectory along the beam axis, thereby emitting photons either left handed or right handed polarised. The difference between the two undulator configurations lies in the emitted on-axis photon beams, where the helical undulator is limited to the fundamental harmonic, while a planar undulator can generate higher harmonics which is attractive to synchrotrons [45]. Helical undulators are attractive to FELs because their emitted fundamental harmonic is more intense on the axis compared to planar undulators, which is crucial for the self-amplified spontaneous emission (SASE) process where only on-axis radiation counts. For beamline users across various research disciplines, there are frequent demands for polarization mode-adjustable undulators to enable improved observation of their samples [46–49]. As of the present review, no literature documenting the development of HTS undulators with adjustable polarization modes has been identified.

HTS undulators typically utilize superconducting coils wound with high-performance HTS wires. Common materials used include the silver-sheathed $Bi_2Sr_2Ca_2Cu_3O_{10+y}$ (Bi-2223) tapes [50], $Bi_2Sr_2CaCu_2O_{8+y}$ (Bi-2212) wires [51], $ReBa_2Cu_3O_{7-\delta}$ (ReBCO, Re = rare earth) tapes (also called coated conductors) [52], $MgB_2$ wires [53] and iron-based superconductors [54]. Among these, ReBCO tapes stand out as particularly promising for practical applications due to their high critical temperature ($T_c$) and critical current density ($J_c$), as illustrated in the summary plot presented in reference [38]. It is important to note that the $J_c$ of these tapes is significantly influenced by the angle between the magnetic field and the surface of the tape. Fig. 2(a-b) depicts magnetic models of HTS planar undulators featuring configurations with two sets of horizontal racetrack (HR) coils and two sets of vertical racetrack (VR) coils, respectively. Both configurations were widely employed in advanced Nb-Ti or $Nb_3Sn$ undulators. Fig. 2(c) illustrates a schematic of an HTS planar undulator utilizing two sets of laser-structured, laminated ReBCO tape stacks [55]. This design ensures controlled paths for the transport current within each tape stack, generating an alternating magnetic field along the beam axis. Fig. 2(d) presents the magnetic model of an HTS helical undulator comprising one pair of helically wound superconducting coils. By applying currents in opposite directions to these pairs, alternating magnetic fields are generated in the *y-z* and *x-z* planes, with a 90-degree phase shift between them.

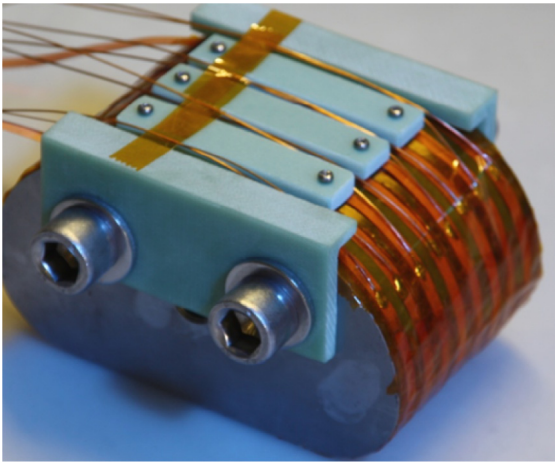

**Fig. 3.** A racetrack ReBCO coil prototype by Boffo et al. Reproduced with permissions from [64].

Magnetized bulk superconductors can be used as pseudo-permanent magnets in HTS undulators. ReBCO bulk superconductors can trap magnetic fields over 17.6 Tesla, making them ideal for compact undulator designs [56,57]. Fig. 2(e) illustrates the magnetic layout of an HTS planar undulator comprising two sets of horizontally arranged ReBCO bulks. This design integrates racetrack coils to generate a uniform dipole field for magnetizing the ReBCO bulks [58,59]. An alternative approach involves constructing an HTS planar undulator using ReBCO bulks with a superconducting solenoid to produce a uniform magnetic field $B_z$. This field magnetizes a series of staggered-array, half-moon-shaped bulk superconductors, as depicted in Fig. 2(f) [60,61]. By rotating each ReBCO bulk continuously by 90 degrees relative to its predecessor and magnetizing the long HTS insert with a uniform magnetic field $B_z$, one obtains a novel type of HTS helical undulator, as depicted in Fig. 2(g) [38].

## 3. Recent progresses

Over the past decade, advances have been made in the development of racetrack HTS coils tailored for undulator applications [62–69]. An example includes the development of a ReBCO coil prototype by Boffo et al, as illustrated in Fig. 3. This prototype was charged with an engineering current density $J_e$ up to 700 kA·mm$^{-2}$, however, magnetic measurements were not conducted. In 2014, Nguyen et al. achieved a milestone with a three-period HTSU utilizing VR ReBCO coils, achieving an undulator field of $B_0 = 0.77$ T, featuring a period of 14 mm and a 3 mm pole gap, operated within a subcooled liquid nitrogen environment at 65 K [66]. Subsequently, Kesgin et al. investigated HTSU prototypes employing non-insulation (NI) or partial-insulation (PI) ReBCO coils, achieving an operational current density up to 2.1 kA·m$^{-2}$ [68]. Their findings highlighted that the PI technique not only reduced charging delays but also maintained the intrinsic self-protection features of NI coils during a quench. In 2024, Garg et al. reported on the fabrication and characterization of an HTS planar undulator utilizing MgB$_2$ coils designed with three periods, featuring a period length of 14.4 mm and a magnetic gap of 6.35 mm, as depicted in Fig. 4 [70]. The inherent asymmetry in the coil configuration resulted in an observed field offset in the undulator, characterized by peaks at +1.19 T and −0.25 T. Numerical simulations indicated that a meter-long MgB$_2$ undulator could achieve an undulator field of $B_0 = 0.85$ T, consistent with theoretical expectations.

In 2009, Prestemon et al. introduced a meander-type HTSU concept utilizing laser-structured ReBCO tapes with predefined current paths, as illustrated in Fig. 2(c) [71,72]. Subsequent experimental validation demonstrated that a single laser-structured tape could generate an alternating magnetic field exceeding 10 mT along the central axis, positioned 2 mm above the tape surface [73]. Further R&D efforts aimed at enhancing the applicability of the meander-type undulator included innovative fabrication techniques such as the JUST winding scheme proposed by Holubek et al. [74] and the soldering method for adjacent ReBCO tapes proposed by Krasch et al. [75]. Numerical simulations indicate that this emerging technique shows potential for achieving substantial undulator field with short period length, though significant prototyping efforts are necessary for practical deployment.

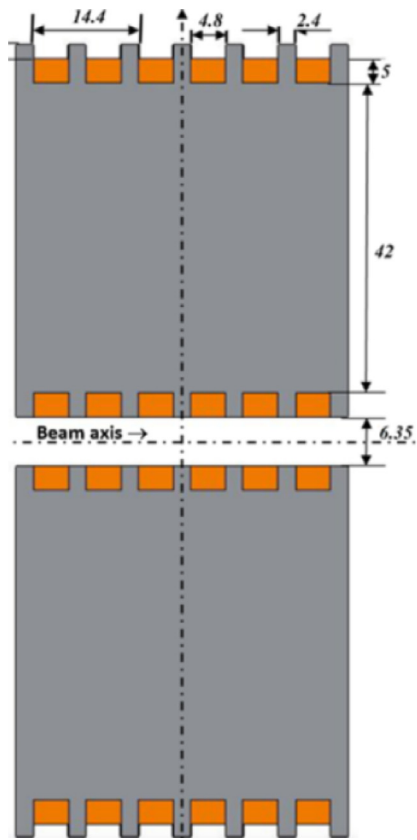

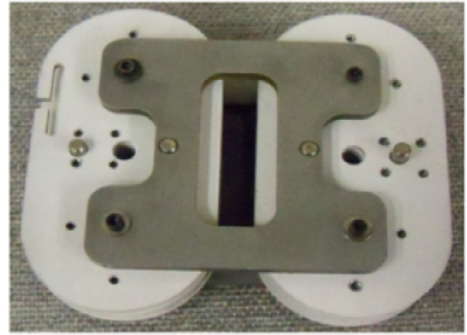

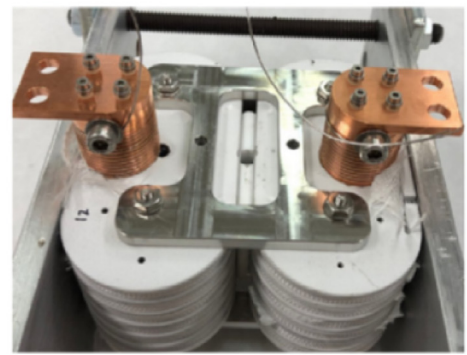

**Fig. 4.** An MgB$_2$ undulator prototype by Garg et al.. Reproduced from [70]. CC BY 4.0.

High-temperature superconducting wires or tapes can be configured in a spiral arrangement to create a helical undulator, depicted in Fig. 2(d). In 2008, Majoros et al. documented the fabrication and testing of a $MgB_2$ helical undulator with a period length of 14 mm and a total length of 250 mm [76]. The prototype, constrained by the filling factor and low $J_c$ of $MgB_2$ round wires, achieved an undulator field of 0.25 T within a 7 mm beam aperture. In 2024, Richter et al. introduced the design and construction of a demonstrator HTS helical undulator comprised of five periods wound using ReBCO tapes [77]. To ensure smooth winding and preserve the tape's $J_c$, the authors developed a specialized mechanical support structure with a generous bending radius to ease the tape's return path. This helical prototype demonstrated stable operation under a charging current of 150 A at 77 K, though detailed characterization of its on-axis magnetic field distribution was not included in the publication.

High-temperature superconducting undulators constructed from bulk superconductors have become increasingly prominent in both the applied superconductivity and accelerator-based photon source communities in recent years. One promising approach involves using a superconducting solenoid to magnetize a series of ReBCO bulk superconductors, thereby generating planar or helical undulator fields, as depicted in Fig. 2(f) and 2(g) respectively. In 2013, Kinjo et al. demonstrated a short bulk HTSU prototype could produce an undulator field of $B_0 = 0.85$ T for $\lambda_u = 10$ mm, within a 4 mm magnetic gap [78]. Subsequently, motivated by the EU-funded CompactLight project initiated in 2018 for designing a compact hard x-ray FEL [79], PSI in collaboration with University of Cambridge undertook research and development efforts focused on short-period bulk HTS undulators. They fabricated a series of 10 mm-period undulator samples using various ReBCO bulk materials and assembly techniques, as illustrated in Fig. 5. Below is a concise summary of their comprehensive fabrication and testing campaign conducted on the undulator samples:

i. The first bulk sample comprised 5 periods with a 6 mm magnetic gap, fabricated using GdBCO/Ag bulks grown by the University of Cambridge, as illustrated in Fig. 5(a). Following field cooling (FC) magnetization from 6 T, the sample produced an undulator field of 0.85 T at 10 K [30]. Notably, during a subsequent test, a higher undulator field was achieved due to delayed quenching, suggesting a potential training effect possibly linked to the movement of half-moon shaped GdBCO disks within copper disks or the relative movement between neighbouring GdBCO-Copper disks under Lorentz forces.

ii. The second bulk sample employed GdBCO/Ag bulks obtained from Cambridge, comprising 10 periods with a 4 mm magnetic gap, illustrated in Fig. 5(b). Upon FC magnetization starting from $B_z = 8$ T, the undulator field showed a gradual increment until a quench occurred at $B_z = 1$ T. Subsequently, as $B_z$ continued to decrease, the undulator field exhibited another rise until a second quench observed at $B_z = -3$ T. This bulk sample achieved an undulator field of 1.29 T at 15 K and 1.54 T at 10 K, respectively.

iii. Based on lessons learned from the initial two bulk samples, a novel shrink-fit assembly technique was developed for industrial bulk samples. This method aimed to apply sufficient pre-stress to half-moon-shaped HTS bulks to suppress potential displacements and withstand significant Lorentz forces during magnetizations. In this technique, each HTS bulk was precisely fitted into a heated copper disk (∼200 °C), followed by assembly into a lengthy aluminum shell also heated to ∼ 200 °C, as illustrated in Fig. 5(c). It was observed that this novel technique could eliminated premature quenches but not the inhomogeneity of the undulator field observed in previous bulk samples. To reduce the peak-to-peak field error of $\sigma B/B_0$, all bulk pieces were sorted and reassembled into a modified aluminum shell support, depicted in Fig. 5(d). Details about the analysis and optimization of the undulator field were described in [80].

iv. The observed inhomogeneity in the undulator field across the samples was attributed to variations in critical current density ($J_c$) among individual bulk superconductors. To address this issue, Calvi et al. proposed a strategy to pre-screen bulk pieces based on their superconducting performance prior to assembly into an undulator [7]. This method was implemented in the fabrication of a new industrial bulk sample utilizing GdBCO bulk superconductors from Nippon Steel (NS) company. The bulk superconductors were housed within newly designed two-piece copper support shells aimed at enhancing thermal conduction, as depicted in Fig. 5(e). Fig. 6(a) illustrates magnetic field scans at 77 K conducted on thirty magnetized, half-moon-shaped NS/GdBCO-Cu pieces. From these scans, twenty pieces demonstrating intermediate trapped field performances were selected, as highlighted in Fig. 6(b). Fig. 6(c-d) present magnetic field measurement results for the selected sample of Fig. 5(e). This optimized bulk sample was magnetized at 10 K and achieved a notable undulator field of $B_0$ = 2.1 T with a 10 mm period and exhibited a low peak-to-peak field error, $\sigma B/B_0$, comparable to that of permanent magnet undulators prior to magnetic shimming [7]. Fig. 7 provides a comparative analysis of achievable $B_0$ values across different undulator technologies. These $B_0$-$\lambda_u$ curves, corresponding to a fixed magnetic gap of 4 mm, were derived through extrapolation from the scaling laws outlined in reference [81].

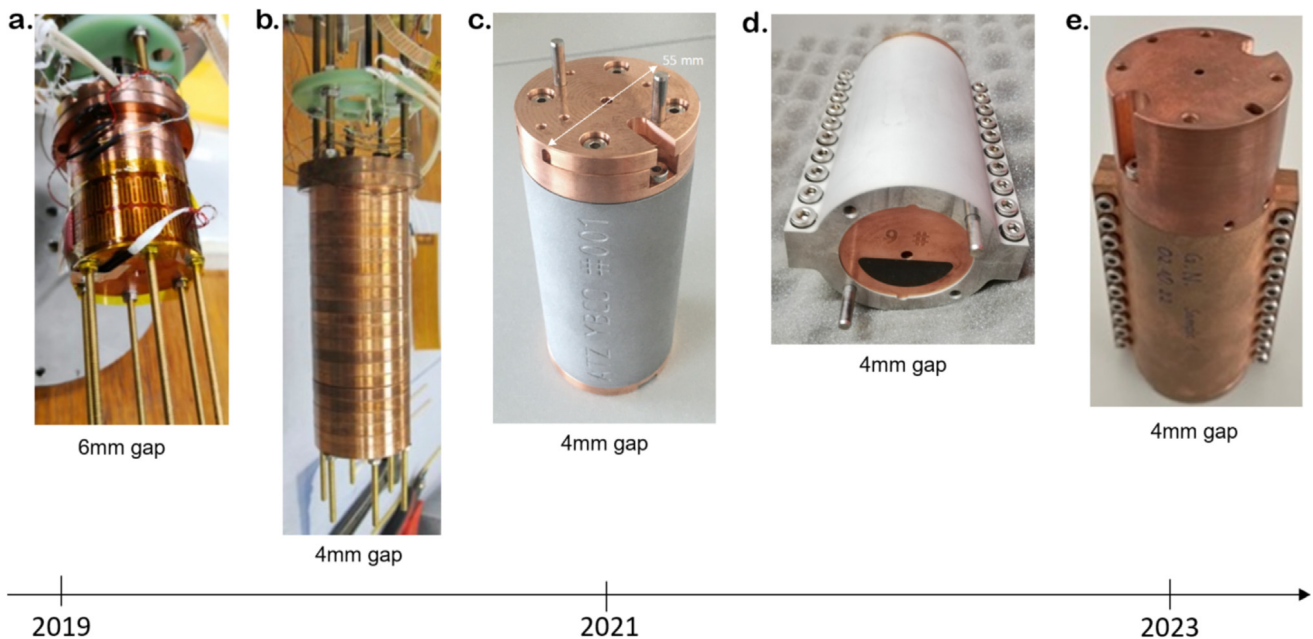

**Fig. 5.** Timeline of the development of HTS planar undulator models comprising of staggered-array ReBCO bulk superconductors at the Paul Scherer Institute.

In 2023, Calvi et al. introduced a novel GdBCO bulk superconducting helical undulator prototype tailored for compact X-ray Free Electron Lasers (FELs), as depicted in Fig. 8(a) [38]. Utilizing GdBCO bulk pieces selected from the previously identified sample shown in Fig. 5(e), the authors arranged each piece at a 90-degree rotation relative to its predecessor, thereby creating a new helical undulator configuration featuring a double staggered-array of bulk superconductors. This 104 mm-long undulator prototype was designed with a period length of 16 mm and a magnetic gap of 4 mm. Upon magnetization with a solenoidal field of 10 Tesla, the prototype produced an on-axis magnetic field exceeding 2.5 T in *y-z* and *x-z* planes, as illustrated in Fig. 8(b-c). It is noteworthy that a helical undulator configuration has the potential to enhance the interaction between electrons and photons [82], leading to earlier saturation of the FEL process and increased power, which in turn reduces the required undulator length.

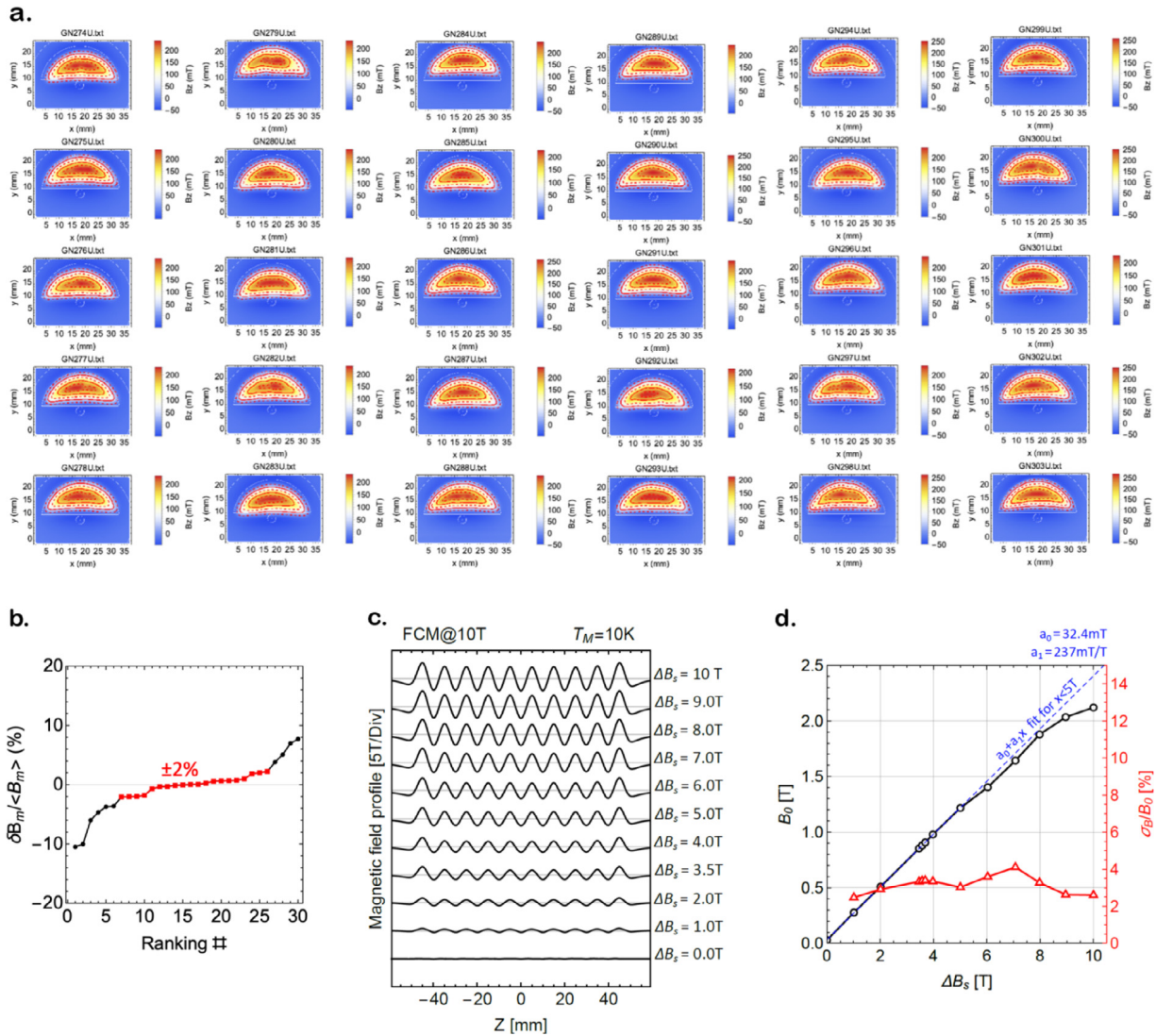

**Fig. 6.** (a) Scanning magnetic field maps of 30 NS-GdBCO/Ag disks magnetized at 77 K. (b) Selection of 20 disks with intermediate trapped field capabilities. (c) Measured magnetic field of a "good" sample comprising of pre-sorted NS-GdBCO/Ag disks. (d) Relation between undulator field $B_0$, the field error $\sigma B/B_0$ and $\Delta B_s$. Reproduced from [7]. CC BY 4.0.

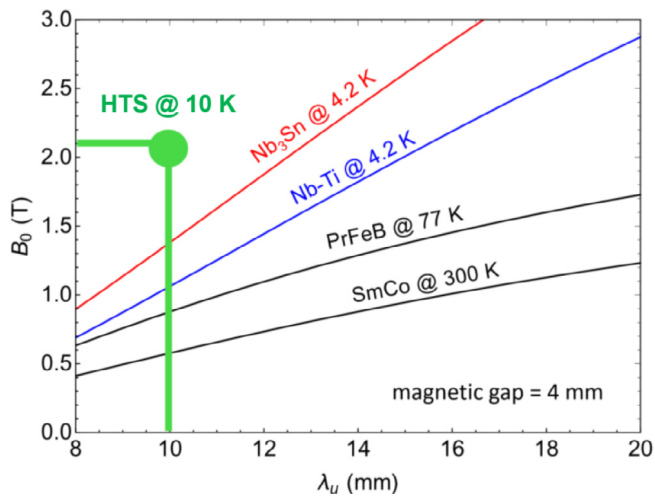

**Fig. 7.** Comparison of the achievable undulator field $B_0$ between using HTS and the other technologies [7,81]. The magnetic gap is fixed at 4 mm.

## 4. Towards practical applications

Paul Scherrer Institute (PSI) has initiated the development of a meter-long bulk HTS undulator for the upcoming I-TOMCAT tomographic imaging beamline at the upgraded Swiss Light Source (SLS 2.0) by late 2026. The envisioned HTS undulator aims to significantly enhance photon brightness by more than an order of magnitude, enabling experiments at photon energies up to 60 keV and expanding the photon spectrum twofold compared to conventional undulator technologies. Fig. 9 provides the flux comparison at 30 m from the source to illuminate a sample of about $1.0 \times 1.0$ mm$^2$ at I-TOMCAT beamline between installing a 10 mm-period HTSU with $B_0 = 2.1$ T and a state-of-the-art 14 mm-period CPMU with $B_0 = 1.3$ T of the same undulator length of 1 m into SLS 2.0 which has an electron beam energy of 2.7 GeV and emittance of 150 pm·rad. Following deliberations, the HTS undulator is designed to be 2.0 m long, with an active length of 1.0 m, a period length of 10.5 mm, a magnetic gap of 4.5 mm, and a peak undulator field $B_0$ of 1.8 T. Fig. 10(a) depicts the appearance of the HTS planar undulator scheduled for installation at the I-TOMCAT beamline at SLS 2.0 [39]. Fig. 10(b) provides a cross-sectional view of the undulator cryostat which houses a 12 T superconducting solenoid wound with Nb$_3$Sn/Cu wires and a variable temperature HTS insert composed of staggered-array ReBCO bulk superconductors. The HTS insert, integrated within the cold solenoid magnet, facilitates temperature control ranging from 4 K to above 100 K using GM cryocoolers and heaters. Fabrication and assembly of hundreds of half-moon-shaped bulk superconductors and ferromagnetic pole pieces have been completed, preparing for the construction of the meter-long HTS insert. The 12 T superconducting solenoid magnet is currently being constructed at Fermi National Accelerator Laboratory (FNAL) and will soon be transported to PSI for integration with the HTS insert [83].

Recently, research and development on a 0.6 m-long, 12 mm-period bulk HTS undulator prototype has been initiated in Shanghai, with the objective of supporting future enhancements to SSRF or SXFEL. This prototype is intended for testing at the Shanghai Soft X-ray Free Electron Laser (SXFEL) facility during its scheduled maintenance in 2025 [42,84]. Fig. 11 illustrates the design of the 12 mm-period bulk HTS undulator prototype, providing a sectional view featuring its two separate vacuum chambers. One chamber is dedicated to housing a 7 T superconducting solenoid, while the other accommodates a variable temperature HTS insert composed of staggered-array ReBCO bulks. Both designs, although slightly reducing magnetic field utilization efficiency, offer the advantage of enabling straightforward assembly and disassembly of the HTS insert without requiring the warm-up of the superconducting solenoid magnet. Such capability streamlines the iterative process necessary for optimizing local magnetic fields and minimizing RMS phase errors. The fabrication, assembly, and testing of the conduction-cooled 7 T superconducting solenoid magnet have been completed. Concurrently, collaborative efforts between the HTSU teams and Shanghai University are progressing towards the development of one-step-grown, half-moon-shaped YBCO bulk superconductors. These advancements hold promise for achieving improved performance characteristics tailored specifically for undulator applications [85].

## 5. Discussion

High-temperature superconducting coils wound with ReBCO tapes have demonstrated the capability to generate static magnetic fields exceeding 20 Tesla in solenoids [86–89]. However, they have not yet surpassed Nb-Ti or Nb$_3$Sn coils in generating high undulator fields.

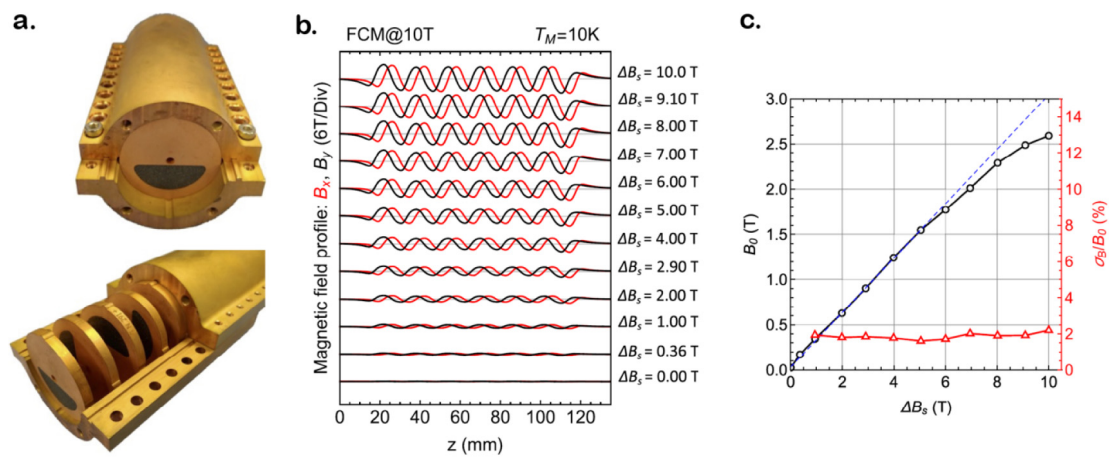

**Fig. 8.** (a) A short helical undulator prototype fabricated based on a novel double staggered-array configuration proposed by Calvi et al. (b) Measured magnetic field profile during FC magnetization. (c) Relation between the amplitude of undulator field, $B_0$ and the reduction in the background magnetic field, $\Delta B_s$. Reproduced from [38]. CC BY 4.0.

This limitation stems from challenges such as detecting and protecting against quenches, managing the screening current induced field (SCIF) effects and even developing suitable winding techniques [90–94]. The non-insulation (NI) winding technique shows potential for quench protection but introduces delays in charging or discharging, necessitating extended stabilization times for the central magnetic field [95–98]. Furthermore, current sharing between turns restricts ReBCO coils from achieving their critical current. Recent efforts have focused on addressing these issues through the partial-insulation (PI) winding technique and reducing AC losses or SCIF effects using striated ReBCO tapes [99–102]. These advancements could minimize charging delays and hysteresis between the undulator field and charging current, as discussed in [5,103]. Moreover, significant advancements have been made in developing high-$J_e$ ReBCO coated conductors, which exhibit superior performance compared to conventional ReBCO tapes [104,105]. The potential commercialization of these enhanced ReBCO tapes holds promise for advancing ReBCO coil-based undulators.

Recent progress in short-period HTS undulators utilizing ReBCO bulk superconductors has garnered attention within the undulator community and broader applications of bulk superconductors. Both planar and helical undulator prototypes have shown markedly higher on-axis magnetic fields compared to Nb-Ti counterparts of equivalent period lengths [7,38]. However, significant challenges remain from the fabrication of short undulator samples to the development of practical bulk HTS undulators suitable for synchrotron and Free Electron Laser (FEL) applications.

i. First, achieving uniform undulator fields requires maintaining consistent temperatures across the lengthy HTS insert, which comprises staggered-array or double-staggered-array ReBCO bulk superconductors. To achieve this thermal stability, cryostat designs for engineering HTSU devices, as developed by PSI and ZJLAB, incorporate dual GM cryocoolers to cool both ends of the HTS insert.
ii. Second, accurate measurement systems for horizontal magnetic fields within vacuum environments are essential for characterizing both local and overall undulator fields in narrow gaps. Techniques such as hall probe scanning and moving wire methods are demanded for high-precision measurements.
iii. Third, to correct local undulator fields and reduce peak-to-peak field errors or RMS phase errors, initial steps involve sorting ReBCO bulks based on their trapped field capabilities. Fine-tuning of local magnetic fields then requires adjusting the local gap of ferromagnetic poles, which involves iterative processes and thermal cycling to achieve the desired undulator field uniformity.
iv. Last but not least, synchrotrons and FELs require undulator devices capable of maintaining stable undulator fields over a period of time for beamline users. To achieve this, preventing flux creep in all magnetized ReBCO bulks is crucial to avoid rapid decay of the induced undulator field. One effective approach involves subcooling the HTS insert to freeze the trapped magnetic flux [7]. Nevertheless, further detailed investigations are essential to address this concern comprehensively.

Undulators capable of switchable polarizations, exemplified by Advanced Planar Polarized Light Emitting (APPLE) undulators, are highly valued by beamline users in advancing scientific discoveries [106]. However, conventional APPLE undulators often feature a long period length $\lambda_u$ constrained by the performance limitations of permanent magnets. To enhance the production of high-energy radiation photons and simplify motor control systems, researchers are exploring the development of superconducting undulators with switchable polar-izations utilizing Nb-Ti coil technology. This includes hybrid planar-tilted configurations and the superconducting arbitrarily polarizing emitter (SCAPE) [107–111]. A promising strategy to advance high-performance superconducting undulators with switchable polarizations involves integrating the SCAPE design with high-$J_e$ ReBCO tapes-based superconducting coils.

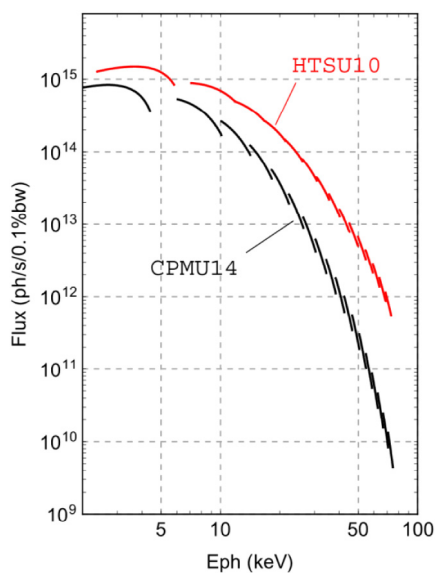

**Fig. 9.** Comparison of the calculated flux at 30 m from the source to illuminate a sample of about $1.0 \times 1.0$ mm$^2$ at I-TOMCAT beamline between installing HTSU10 with $B_0 = 2.1$ T and CPMU14 with $B_0 = 1.3$ T of the same undulator length of 1 m into SLS 2.0 which has an electron beam energy of 2.7 GeV and emittance of 150 pm·rad.

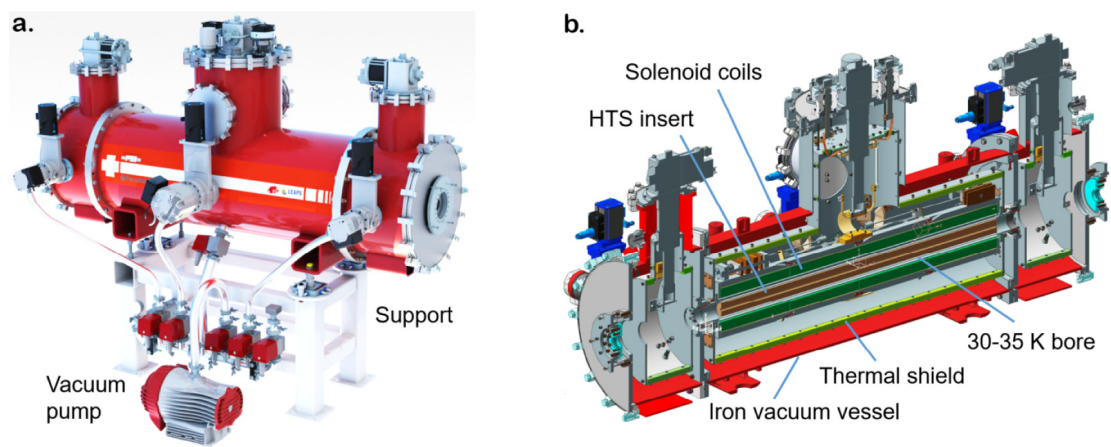

**Fig. 10.** (a) A 1 m-long, 10.5 mm-period HTS planar undulator designed for the I-TOMCAT beam line at SLS2.0. (b) Half section view of the cryostat which houses the 12 T superconducting solenoid and the variable temperature HTS insert.

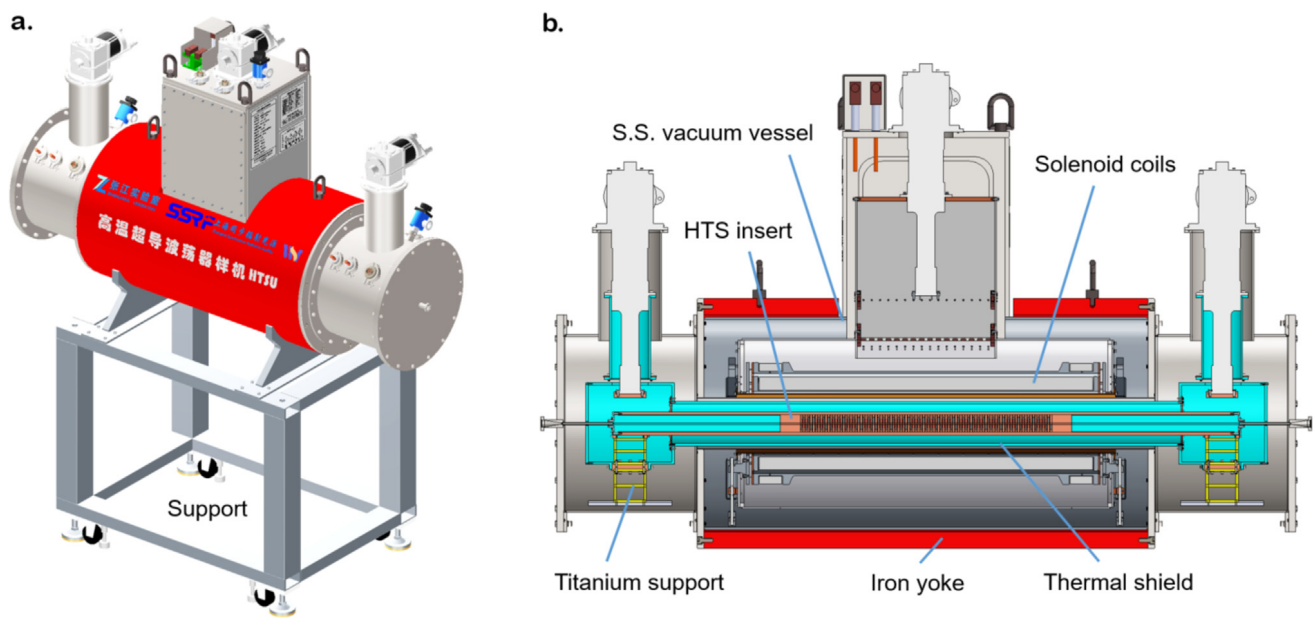

**Fig. 11.** (a) A 0.6 m-long, 12 mm-period HTS planar undulator under construction in Shanghai. (b) Half section view of the cryostat with two independent vacuum chambers, one for housing the 7 T superconducting solenoid and the other for housing the variable temperature HTS insert.

## 6. Conclusions

In this work, we have reviewed the working principles of and recent advancements in various types of HTS undulators utilizing ReBCO tapes and bulks. Significant efforts have been focused on developing ReBCO coil-based HTS undulators, however, progress has been impeded by inherent challenges such as quench protections and screening current induced field effects. Nonetheless, progress in enhancing engineering critical current density and employing striated ReBCO tapes alongside partial-insulation winding techniques holds potential for the evolution of HTS coil-based undulators. Concurrently, breakthroughs have been achieved in research on both planar and helical HTS undulators utilizing ReBCO bulk superconductors, demonstrating undulator fields with at least doubled performance compared to Nb-Ti superconducting undulators of equivalent period lengths and magnetic gaps. Despite challenges such as precise on-axis magnetic field measurement, local field corrections, and managing flux creep-induced field decay, global undulator teams are advancing towards developing engineering bulk HTSU devices for practical applications.

## CRediT authorship contribution statement

**Zhuangwei Chen:** Writing – original draft. **Marco Calvi:** Writing – review & editing. **John Durrell:** Writing – review & editing. **Cristian Boffo:** Writing – review & editing. **Dabin Wei:** Writing – review & editing. **Kai Zhang:** Writing – review & editing, Writing – original draft, Supervision. **Zhentang Zhao:** Writing – review & editing, Supervision.

## Data availability

Data will be made available on request.

## Declaration of competing interest

The authors declare that they have no known competing financial interests or personal relationships that could have appeared to influence the work reported in this paper.

## Acknowledgement

This work is supported by the Chinese National Overseas Yang Talent program, the Swiss Accelerator Research and Technology (CHART) program and the European Union's Horizon 2020 research and innovation program under grant agreement no. 101004728 (LEAPS-INNOV). This work has been partially funded by Swiss LOS Lottery Fund of Kanton Aargau. This work has been authored by Fermi Research Alliance, LLC under Contract No. DE-AC02-07CH11359 with the U.S. Department of Energy, Office of Science, Office of High Energy Physics.